\documentclass[%
 reprint,
amsmath,amssymb,
aps,
pra,
longbibliography,
floatfix,
]{revtex4-2}

\usepackage{graphicx}
\usepackage{dcolumn}
\usepackage{bm}
\usepackage{siunitx}
\usepackage{physics}
\usepackage{simpler-wick}
\usepackage{graphicx}
\usepackage{enumerate}
\usepackage[hidelinks]{hyperref}
\usepackage[many]{tcolorbox}    	
\usepackage[english]{babel}
\newtheorem{theorem}{Theorem}

\AtBeginDocument{\RenewCommandCopy\qty\SI}
\begin{document}



\title{Decoherence-Free Subspaces Cannot Protect Against\\Spontaneous Wave Function Collapses}

\author{Alfred Li}
\email{ali2@caltech.edu}
\affiliation{Department of Physics, Princeton University, Princeton, NJ 08544, USA}
\affiliation{The Division of Physics, Mathematics and Astronomy, California Institute of Technology, Pasadena, CA 91125, USA}

\author{Herschel A. Rabitz}
\affiliation{Department of Chemistry, Princeton University, Princeton, NJ 08544, USA}

\author{Benjamin Lienhard}
\email{blienhard@princeton.edu}
\affiliation{Department of Chemistry, Princeton University, Princeton, NJ 08544, USA}
\affiliation{Department of Electrical and Computer Engineering, Princeton University, Princeton, NJ 08544, USA}

\date{\today}

\begin{abstract}
Efficacious quantum information processing relies on extended coherence and precise control. Investigating the limitations surrounding quantum processors is vital for their advancement. In their operation, one challenge is inadvertent wave function collapse. Decoherence-free subspaces, theoretically capable of mitigating specific non-unitary dynamics, present a promising avenue. However, our study unveils their inability to safeguard against spontaneous wave function collapse. Thus, the spontaneous collapse rate becomes a critical limiting factor for quantum systems' physical coherence, restricting the options for maintaining coherence to active error correction.
\end{abstract}

\maketitle





\textit{Introduction.---}Quantum information processing hinges on the quantum system's ability to maintain coherence and preserve the integrity of quantum information. However, any realistic system will inevitably couple and interact with the environment, thus leading to effectively non-unitary dynamics known as decoherence. Dephasing and generalized amplitude damping capture the prototypical forms of decoherence~\cite{Nielsen2011_QuantumComputationQuantum}. 


Suppose a quantum system initially exists in an unentangled state. As time elapses, the quantum system can get entangled with its environment, leading to the spatial expansion of its wave function. The entanglement process typically manifests itself through phase accumulation. This ``dephasing'' tendency can often be mitigated by inducing a reversal in the sign of the phase accumulation~\cite{Hahn1950_SpinEchoesa}. However, the wave function can collapse spontaneously, rendering the phase accumulation irreversible. Additionally, these collapses can project the quantum system of interest onto one of its eigenstates, a process described by generalized amplitude damping. If the projection of the quantum system of interest onto an eigenstate is associated with energy loss, the process is equivalent to spontaneous emission~\cite{EinsteinPhysZ1917SPE}. Typically, the energy loss, and thereby the encoded information, is irretrievable, as is the loss of quantum information during wave function collapses. Hence, mastering the understanding and mitigation of undesired wave function collapses is crucial in quantum information processing.

The statistical transition rates between quantum states are captured by Fermi's Golden Rule~\cite{Fermi1950FGR}. The transition rates depend on the coupling between the initial and final states as well as the density of states at the energy of the final state. 
The design of quantum information processors primarily focuses on reducing the system-bath coupling rates. Quantum control---the control of the quantum system through modulated electromagnetic fields---can further increase the coherence of quantum systems. For example, the environment of superconducting qubits comprises a host of unknown quantum systems, often categorized and referred to as two-level systems due to imperfect host materials. The transition frequencies of these spurious atomic tunneling defects can be tuned away from the qubit's resonance frequency through an applied DC-electric field, reducing the system-bath coupling rates and enhancing the qubit's coherence time~\cite{LisenfeldNPJQ2023TLS}. Protocols to manipulate the coupling rates through quantum control have also been demonstrated for other quantum processor modalities such as quantum dots~\cite{jin_natnano2014_quantumdots}. 

A few approaches to alter the density of states have been suggested and demonstrated experimentally. For instance, unpaired electrons, so-called quasiparticles, can diminish the lifetime of superconducting qubits. A qubit's lifetime can be increased through a sequence of control pulses, effectively lowering the number of quasiparticles~\cite{gustavsson_science2016_relaxation_SCQ}.

In addition to minimizing the spontaneous transition events, preventing energy loss or gain is possible under some circumstances~\cite{AlberPRL2001SPE,AlberPRA2003SPE,KhodjastehPRL2002SPE}. Monitoring specific characteristics of a quantum system can reveal a tell-tale signal of these discrete and random events. Using a superconducting artificial three-level atom, quantum jumps from the ground state to an excited state have been tracked and predicted by monitoring an auxiliary energy level coupled to the ground state~\cite{MinevNature2019QuantumJump}. Consequently, the evolution of each jump is continuous, coherent, and deterministic, allowing for real-time monitoring and feedback to catch and reverse incomplete quantum jumps.

\begin{figure*}[t]
\includegraphics[width=\textwidth]{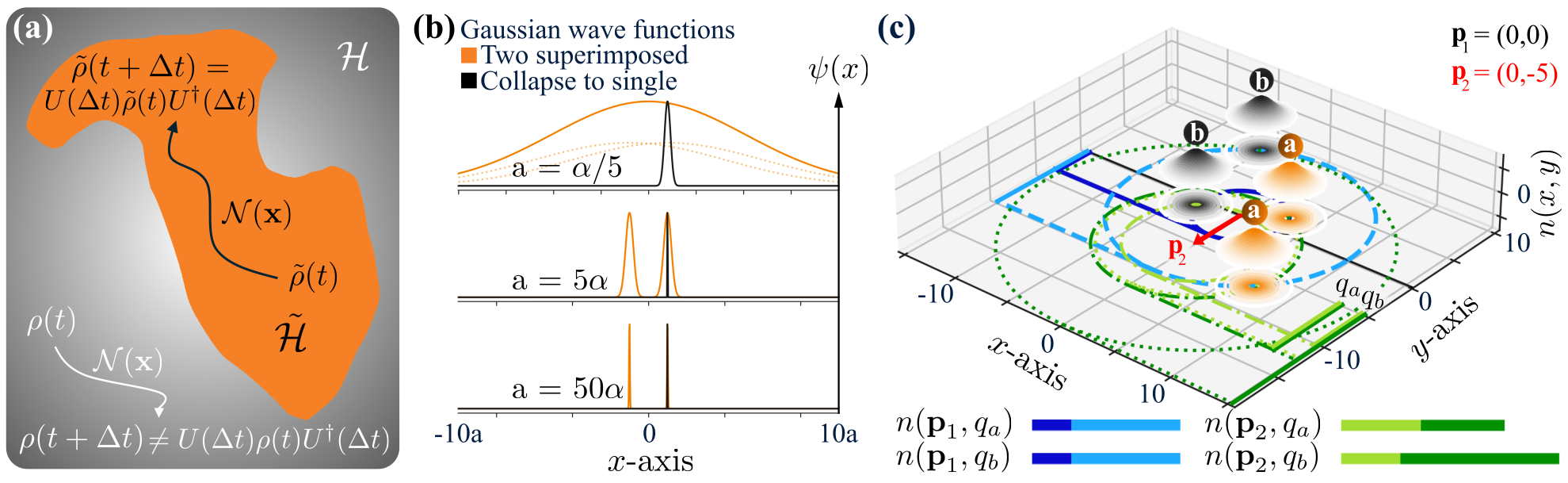}
\caption{\textbf{Decoherence-Free Subspaces and Continuous Spontaneous Localization.} \textbf{(a)} Graphical illustration of a decoherence-free subspace (DFS) \(\tilde{\mathcal{H}}\) within Hilbert space \(\mathcal{H}\). States within this subspace undergo evolution solely through unitary dynamics \(U(t)=\exp{-i\mathcal{H}_S t/\hbar}\) and system Hamiltonian \(\mathcal{H}_{S}\), remaining impervious to the erroneous processes induced by environmental interactions, captured by \(\mathcal{N}(\textbf{x})\). \textbf{(b)} Continuous Spontaneous Localization (CSL) mathematically models the collapse of wave functions. For a wave function \(\psi(x)\) comprising two superimposed Gaussian functions with spatial separation \(2a\), the collapse probability and localization to a single Gaussian function increase with rising \(a\). \textbf{(c)} The eigenvalues \(n(\textbf{p}_1, q_a)\) (indicated in orange and subscript \(a\)) and \(n(\textbf{p}_1, q_b)\) (colored in black and denoted with subscript \(b\)) with Gaussian functions centered at positions \(q_a\) and \(q_b\) are equivalent if evaluated at position \(\textbf{p}_1\). The eigenvalues differ if evaluated at \(\textbf{p}_2\).}
\label{Fig}
\end{figure*}

An alternative pathway to design active control of wave function collapses, preemptively or reactively, is to process information using quantum states that are inherently robust to it. A set of states forming a codespace and robust to specific decoherence effects is known as a noiseless subsystem, or decoherence-free subspace (DFS)~\cite{ZanardiMPLB1997ErrorAvoidingCodes, ZanardiPRA1998DFS, Lidar_2003,ShabaniPRA2005DFS,BlumePRL2008DFS,LidarACP2014DFS}. The realizability of DFSs has been verified, both theoretically and experimentally, in different quantum systems ranging from optical systems~\cite{KwiatScience2000DFS} to quantum dots~\cite{FriesenNatCom2017QDDFS} and nuclear spins~\cite{ReisererPRX2016NVDFS, BartlingPRX2022NVDFS}. While experimental demonstrations of DFS spontaneous emission cancellation~\cite{MarzlinCJP2007DFSSPE} leveraging destructive interference have been reported~\cite{XiaPRL1996EXP}---their results have faced scrutiny and questioning~\cite{LiPRL2000EXPCrit}---it remains an open question whether DFS exist that can protect against wave function collapses and, thus, spontaneous emission. The concept of a DFS is graphically depicted in Fig.~\ref{Fig}\textbf{(a)}.

Modeling the process of spontaneous emission and its underlying physical cause has been a subject of research for nearly 90 years~\cite{weisskopf_natwissen1935_spontaneousemission,ginzburg_sssr1939_spontaneousemission, fain_ncb1982_vacuumfluctuations}. Collapse models are a class of methodologies to capture the effects of wave function collapses phenomenologically and, consequently, of spontaneous emission without defining the underlying physical causes~\cite{BassiEntropy2023CSLReview, CarlessoNatPhys2022CollapseReview, AdlerScience2009CollapseModels}. Collapse models represent a promising attempt to address the ``measurement problem'' of the Copenhagen interpretation of quantum mechanics, the primary motivation for their conception~\cite{heisenberg_1952_measurementproblem}. 

Collapse models amend the Schrödinger equation by introducing nonlinear and stochastic ``noise'' terms~\cite{Ghirardi_PRD1986_GRW, Ghirardi_PRA1990_CSL,Bassi_2003, BassiRevModPhys2013CollapseModels}. The modifications have to be such that the quantum system's properties are preserved microscopically, and hence, the predictions of the original and amended Schrödinger equation agree. However, at the macroscopic level, the additional terms must lead to a localized wave function collapse, resulting in a description in accord with classical physics. Among the various collapse models, the continuous spontaneous localization (CSL) model is the most promising candidate and, thus, the most extensively researched and applied~\cite{Ghirardi_PRA1990_CSL}. Fig.~\ref{Fig}\textbf{(b)} contains a graphical illustration of the working principles of CSL. Interestingly, the generation rate of the aforementioned quasiparticles affecting the lifetime of superconducting qubits may depend on CSL~\cite{VischiPRB2022CSLSCQ}.


Analogous to the noise underlying collapse models responsible for the wave function collapse, the common form of spontaneous emission is caused by noise known as vacuum fluctuations~\cite{Murch_Weber_Beck_Ginossar_Siddiqi_2013}. In collapse models, the noise term is typically modeled as white, resembling background fluctuations with zero mean and no spatial or temporal correlation, which sporadically induces wave function collapses. Similarly, in electromagnetic fields, vacuum fluctuations---also known as quantum fluctuations---are defined as the finite standard deviation of the electric field operator relative to the zero-photon Fock state, which also has a zero expectation value. The magnitude of the vacuum fluctuations is equivalent to the ``vacuum electric field amplitude,'' which is spatially and temporally uncorrelated. The similarities between spontaneous emission and collapse models make distinguishing the origin of wave function collapses challenging.


In this article, we investigate the feasibility of constructing a DFS to mitigate unintended wave function collapses, such as those arising from spontaneous emission events. Our findings demonstrate that engineering such a subspace is impossible for wave function collapses captured by collapse models using radial basis functions. Consequently, addressing the adverse impacts of wave function collapses necessitates implementing strategies involving active corrective measures and may limit the protective reach of DFS to continuous dephasing mechanisms. Moreover, in the Appendix, 
we explore the potential consequences of spontaneous collapses on physical qubits embedded in or imprinted on solid-state materials. Our analysis reveals that the coherence limit imposed by wave function collapses can be on the order of seconds and, thus, may already be approaching.

\vspace{5mm}\textit{Decoherence-Free Subspaces.---}A DFS, denoted as \(\tilde{\mathcal{H}}\), is a subspace of a Hilbert space \(\mathcal{H}\) associated with a quantum system. Notably, the states within \(\tilde{\mathcal{H}}\) remain unaffected by system-environment couplings during time evolution, ensuring that they evolve unitarily.

Generally, the master equation for the system density matrix \(\rho(t)\) and Linbladian \(\mathcal{L}\) has the form
\begin{equation}\label{EQ:Lindblad}
    \begin{split}
        &\frac{d}{dt}\rho(t) = -\frac{i}{\hbar}[\mathcal{H}_{S},\rho(t)] + \mathcal{L}[\rho(t)]\\
        &\mathcal{L}[\rho(t)] = \frac{1}{2}\sum_{\nu,\mu = 1}^{M}a_{\nu\mu}\left([F_{\nu},\rho(t)F^{\dagger}_{\mu}]+[F_{\nu}\rho(t),F^{\dagger}_{\mu}]\right),
    \end{split}
\end{equation}
where \(\mathcal{H}_{S}\) represents the system Hamiltonian describing the unitary dynamics while \(F_{\nu}\) denotes non-unitary jump operators that capture decoherence effects. The coefficients \(a_{\nu\mu}\) are elements of a positive semi-definite matrix that comprise information about the bath and depend on the microscopic model of the decoherence effects. 

The condition for a DFS to exist, then, follows as \(\mathcal{L}[\rho(t)] = 0\). Without any assumptions on the bath coefficients \(a_{\nu\mu}\) or density matrix \(\rho(t)\), the following Theorem~\cite{Lidar_2003} succinctly defines the necessary conditions for a DFS to exist:
\begin{theorem}\label{TH:DFS}
A necessary and sufficient condition for a subspace \(\tilde{\mathcal{H}}\) spanned by \(\{\tilde{\ket{k}}\}_{k = 1}^{N}\) to be decoherence-free is that all \(\tilde{\ket{k}}\) are degenerate eigenstates of each Lindblad operator \(\{F_{\nu}\}_{\nu = 1}^{M}\), i.e., 
\begin{equation}
\label{EQ:DFSCOND}
    F_{\nu}\tilde{\ket{k}} = c_{\nu}\tilde{\ket{k}}, \hspace{0.3cm}\forall\hspace{0.1cm}\nu,\tilde{k}.
\end{equation}
\end{theorem}

Determining whether the coupling between the environment and system allows for a DFS to exist then simplifies to finding the eigenstates of the jump operators \(F_{\nu}\) and checking whether Eq.~(\ref{EQ:DFSCOND}) is satisfied. Additional details on Theorem~\ref{TH:DFS} and the concept of DFS can be found in the Appendix.

\vspace{5mm}\textit{Continuous Spontaneous Localization.---}Physical collapse models aim to address the measurement problem by mathematically modifying the standard quantum mechanical framework. 
The most promising and best-studied collapse model is continuous spontaneous localization (CSL)~\cite{Ghirardi_PRA1990_CSL}. Additional information can be found in the Appendix 
and a recent review~\cite{BassiEntropy2023CSLReview}. 


A generic characteristic of CSL is the ambiguity behind the ``preferred basis''~\cite{Bassi_2003}, which is the set of operators that will physically characterize the state vector reductions. 
Following Reference~\cite{Bassi_2003}, we will consider the simple choice of using the local number operator as our preferred basis (the generalization to the mass-proportional CSL is discussed in the Appendix). 
We define the creation and annihilation operators for a particle of spin-component \(s\) at position \(\textbf{y}\) as \(a^{\dagger}(\textbf{y},s)\) and \(a(\textbf{y},s)\). The spatially-dependent number density operator is defined as
\begin{equation}
\label{number density op}
    \mathcal{N}(\textbf{x}) = \sum_{s}\int d\textbf{y}g(\textbf{y}-\textbf{x})a^{\dagger}(\textbf{y},s)a(\textbf{y},s) 
\end{equation}
with a spherically symmetric, positive real function \(g(\textbf{x})\) peaked around \(\textbf{x} = 0\). The operators \(\mathcal{N}(\textbf{x})\) are Hermitian and commute with each other. \(g(\textbf{x})\) is normalized according to \(\int d\textbf{x}\hspace{0.05cm}g(\textbf{x}) = 1\) such that \(\int d\textbf{x}\hspace{0.05cm}\mathcal{N}(\textbf{x}) = N \) where \(N\) denotes the total number of particles in the system. \(g(\textbf{x})\) is typically a Gaussian function expressed as
\begin{equation}
    g(\textbf{x}) = \left(\frac{\alpha}{2\pi}\right)^{3/2}\exp{-\frac{\alpha}{2}(\textbf{x})^2},
\end{equation}
where \(\alpha^{-3/2}\) represents the volume over which the integral in \(\mathcal{N}(\textbf{x})\) is defined. Specifically, \(\alpha\) defines a characteristic distance \(r_c=1/\sqrt{\alpha}\) beyond which localization processes strengthen. The exact value of the constant parameter \(1/\sqrt{\alpha}\) is still subject to research and is expected to be on the order of \SI{100}{\nano \meter}~\cite{ZhengPRR2020TestCSL}, but may range up to the \si{\milli\meter}-scale~\cite{CarlessoNatPhys2022CollapseReview}. 

Physically, the choice of \(\mathcal{N}(\textbf{x})\) as the preferred basis means that the number of particles within any given volume of size \(\alpha^{-3/2}\) will always be definite. \(g(\textbf{x})\) is a probability measure for how likely a state vector reduction is relative to each particle's position. Consequently, wave functions representing configurations with more particles clustered around \(\textbf{x}\) will have elevated localization probabilities. 

The eigenvectors of \(\mathcal{N}(\textbf{x})\) are
\begin{equation}
\label{EQ:Eigen}
    \ket{q,s} = \mathfrak{N}a^{\dagger}(\textbf{q}_{1},s_{1})a^{\dagger}(\textbf{q}_{2},s_{2})...a^{\dagger}(\textbf{q}_{n},s_{n})\ket{0}
\end{equation}
with a normalization constant \(\mathfrak{N}\) and the vacuum state \(\ket{0}\) comprising \(0\) particles. 
The corresponding eigenvalues are
\begin{equation}
    \begin{split}
    n(\textbf{x}) &= \sum_{i =1}^{n}g(\textbf{q}_{i}-\textbf{x})\\
    &= \left(\frac{\alpha}{2\pi}\right)^{3/2}\sum_{i}\exp{-\frac{\alpha}{2}(\textbf{q}_{i} -\textbf{x})^2}.
    \end{split}
\end{equation}

The modified time-dependent Schrödinger equation 
can be transformed into a Lindblad-type representation, similar to Eq.~(\ref{EQ:Lindblad}), with
\begin{equation}\label{EQ:CSL}
    \begin{split}
    \mathcal{L}[\rho(t)] =&\phantom{+} \frac{\gamma}{2}\int d\textbf{x}\left[\mathcal{N}(\textbf{x}),\rho(t)\mathcal{N}(\textbf{x})\right]\\
    &+ \frac{\gamma}{2}\int d\textbf{x}\left[\mathcal{N}(\textbf{x})\rho(t),\mathcal{N}(\textbf{x})\right],
    \end{split}
\end{equation}
where \(\gamma\) is a positive damping rate associated with the ``hitting process'' initiating the collapse. Similar to \(\alpha\), the value of \(\gamma\) is not rigorously determined yet and has been argued on physical grounds to be around \SI{e-9}{\cubic\nano\meter\per\second} so that the mean localization rate \(\lambda\) (the relation is \(\lambda = \gamma(\alpha/(4\pi))^{3/2}\)) of an individual particle is around \SI{e-17}{\per\second}, a sufficiently small number so that the microscopic dynamics are rarely perturbed, but large enough so that the macroscopic reduction rate is non-trivial~\cite{Bassi_2003}. The unexplored parameter range of \(\lambda\) is between \SI{e-10}{\per\second} to \SI{e-20}{\per\second}~\cite{CarlessoNatPhys2022CollapseReview}. There are also experimental studies underway to investigate CSL with a larger noise coupling to understand whether reduction occurs at the nano-scale~\cite{CarlessoNatPhys2022CollapseReview}. A discussion on the parameter range and the impact on quantum technologies can be found in the Appendix.

\vspace{5mm}\textit{Decoherence-Free Subspaces and Continuous Spontaneous Localization.---}The objective is to explore the existence of a DFS in a system affected by the collapse of wave functions captured by CSL. 
A comparison of Eq.~(\ref{EQ:CSL}) and Eq.~(\ref{EQ:Lindblad}) leads to the identification \(F_{\nu} \leftrightarrow \mathcal{N}(\textbf{x})\). For a DFS to exist, a set of \(\tilde{\ket{k}}\) needs to exist such that
\begin{equation}
\label{EQ:DFSCondition}
    \mathcal{N}(\textbf{x})\tilde{\ket{k}} = c(\textbf{x})\tilde{\ket{k}} \hspace{0.3cm} \forall\hspace{0.1cm}\textbf{x},\tilde{k},
\end{equation}
where \(\textbf{x}\) can range over the entire space in which the system is located. The eigenstates \(\tilde{\ket{k}}\) are represented by \(\ket{q_{i},s}\) defined in Eq.~(\ref{EQ:Eigen}). Without loss of generality, suppose the system has \(N\) excitations, and we place no restraints on the spin. Note that two spatially co-located states \(\ket{q,s_a}\) and \(\ket{q,s_b}\) are indistinguishable for bosons and become entangled for fermions, a consequence of the Pauli exclusion principle. Thus, a DFS formed using the spin degree of freedom can be ruled out. The locations \(\{\textbf{q}_{i,1},\textbf{q}_{i,2},...,\textbf{q}_{i,N}\}\) comprised in \(q_{i}\) are distinct. Using Eq.~(\ref{EQ:DFSCondition}), it follows
\begin{equation}
\label{CSL+EQ:DFSCOND}
    \mathcal{N}(\textbf{x})\ket{q_{i},s} = \left(\frac{\alpha}{2\pi}\right)^{3/2}\sum_{j=1}^{N} e^{-\frac{\alpha}{2}(\textbf{q}_{i,j} -\textbf{x})^2}\ket{q_{i},s}
\end{equation}

\newpage
\onecolumngrid
\twocolumngrid
\noindent for all positions \(\textbf{x}\) in the system and for all \(q_{i}\) (i.e., different allocations of the \(N\) excitations) in the DFS. 

To simplify the problem, we focus first on a single \(\mathcal{N}(\textbf{x})\) and study the set of \(\ket{q_{i},s}\) that are degenerate. The easiest construction involves picking an arbitrary distance \(\textbf{r}\) from a specific position \(\textbf{p}\) and constructing a \((d-1)\)-sphere of radius \(\textbf{r}\) around \(\textbf{p}\), where \(d\) is the spatial dimension of the physical system. Any set of \(N\) points on this sphere can be sourced to place the particles, meaning there are an infinite number of degenerate \(\ket{q_{i},s}\) for a given \(\textbf{r}\) and \(\textbf{p}\). This construction takes advantage of the fact that the Gaussian functions are spherically symmetric. 

Since \(\textbf{r}\) can be of arbitrary length, an infinite number of these degenerate ``spheres'' exist. Fig.~\ref{Fig}\textbf{(c)} shows this geometrical approach to form degenerate eigenvectors.

The geometric approach illustrates that, for any single point in space, degenerate eigenstates with distinct sets of particle locations \(\textbf{q}_i\) can be constructed. We seek to verify if one can construct a set of eigenstates that are degenerate at each point in space. The logic of the proof goes as follows: suppose there exists a DFS robust to wave function collapses modeled by CSL, then evaluate the restrictions on \(\textbf{q}_{i}\). In the Appendix, 
we show that it is impossible to satisfy Eq.~(\ref{CSL+EQ:DFSCOND}) for more than one point in space at a time. The findings are summarized in the subsequent Theorem~\ref{TH:DFSCSL}. 

\begin{theorem}\label{TH:DFSCSL}
Let the eigenvalues \(n(\textbf{x}, q)\) of operator \(\mathcal{N}(\textbf{x})\) be represented by a sum of positive radial basis functions. Then, two eigenvectors \(\ket{q_{1},s}\) and \(\ket{q_{2},s}\) with \(q_{1} \neq q_{2}\) can only be degenerate for a unique choice of \(\textbf{x}=\textbf{p}\).

\noindent Consequently, the following eigenvalue equation holds
\begin{equation}
    \mathcal{N}(\textbf{p})\ket{q_i,s} = n(\textbf{p})\ket{q_i,s} \hspace{0.3cm} \forall\hspace{0.1cm}q_i,
\end{equation}
but is in general
\begin{equation}
    \mathcal{N}(\textbf{x})\ket{q_i,s} \neq n(\textbf{x})\ket{q_i,s} \hspace{0.3cm} \forall\hspace{0.1cm}\textbf{x}\neq \textbf{p},\,q_i,
\end{equation}
with \(i\in\{1, \dots, m\},\hspace{0.1cm}m\geq 2\).
\end{theorem}

\vspace{5mm}\textit{Discussion.---}Wave function collapses predicted by CSL may pose an upper rapidly-approaching coherence limit for quantum technologies utilizing solid-state materials. Quantum error correction plays a key role in mitigating the impact of such spontaneous wave function collapses. Errors induced by wave function collapse can be detected and corrected by encoding quantum information in carefully designed and executed quantum error-correcting protocols~\cite{CampbellNature2017QEC}. 

Reducing the substantial effort required for quantum error correction~\cite{AcharyaNature2023QEC} is increasingly crucial to harness the practical utility of quantum processors at an accelerated pace. To alleviate the effort on the logical level within the quantum processor stack through quantum error correction, passive protective means on the physical level have been proposed~\cite{LidarPRL1999DFSQEC, SchlosshauerPR2019QD, GuoSB2021QEC}. DFSs stand out as one of the most promising methodologies for passive safeguarding quantum information. These specialized subspaces, nestled within the overall Hilbert space of the system, possess a distinctive quality---immunity to decoherence. DFSs exploit the symmetries in the coupling between qubits and the noise. In response to the challenge of limited symmetries in numerous quantum systems, a recent study successfully showcased experimentally the generation of symmetries induced by dynamical decoupling~\cite{quiroz2024dynamically}.

We explored the possibility of creating a DFS to alleviate unintended wave function collapses. Our research reveals the impossibility of constructing such a subspace for wave function collapses using models based on radial basis functions. While there is no proof yet that the mathematical framework of collapse models reflects reality~\cite{CarlessoPRD2016CSLEXP,GasbarriComPhy2021Collapse,ArnquistPRL2022CollapseModel}, they persist as robust contenders for addressing the measurement problem~\cite{hance_jpc2022_measurementproblem}. 

Our research findings resonate with a prior theoretical investigation~\cite{KarasikPRA2007MultiparticleDFS} focused on spontaneous emission dynamics. The study posits the existence of DFS specifically for particles co-located in three-dimensional environments. This focus may constrain the applicability of the findings to scenarios in lower-dimensional spaces, where it has been suggested that DFS could mitigate the effects of spontaneous emission~\cite{RossiPRB1999OneDDFS}. In contrast, our study unveils the general nonexistence of DFS offering resilience against wave function collapses for spatially dispersed quantum systems.


In conclusion, while decoherence-free subspaces demonstrate effectiveness in mitigating continuous dephasing, their capacity to compensate for wave function collapse and their effects is questionable, prompting scrutiny of the terminology `decoherence-free.' Consequently, countering the adverse effects of wave function collapses necessitates the implementation of active quantum control techniques, such as quantum error correction (QEC) protocols. However, traditional QEC protocols often rely on entanglement~\cite{bravyi2024entanglement}, potentially requiring the avoidance of specific codes to minimize entanglement and the consequential increase in the wave function collapse rate. Moreover, even with QEC protocols optimized for minimal entanglement, performance degradation may occur with an increasing number of qubits in quantum processors, particularly for tasks reliant on intricate entangled states due to heightened wave function collapse rates. Thus, a judicious balance between error correction techniques and computational algorithms, considering their entanglement requirements, may be crucial for optimizing quantum processor performance. In summary, wave function collapse rates, as predicted by continuous spontaneous localization, could significantly limit the capabilities of large-scale quantum processors.

\vspace{5mm}
H.A.R. acknowledges funding from the DOE (Contract DE-FG02-02ER15344). B.L. is supported by the Swiss National Science Foundation (Postdoc.Mobility Fellowship grant \#P500PT\_211060). The authors thank Stephen L. Adler, Lajos Di\'{o}si, and Paolo Zanardi for helpful discussions and feedback. 

\appendix

\vspace{5mm}\textit{Appendix A: Decoherence-Free Subspace (DFS).---}A DFS is a subspace within a Hilbert space with unitary quantum dynamics influenced by potentially non-Hermitian operators, \(F_\nu\). These operators typically describe the interactions between a quantum system of interest and the surrounding bath. A DFS exploits the symmetries in the system-bath coupling such that states comprised by the subspace remain within the subspace upon being acted on by operator \(F_\nu\).

A common example to illustrate the working principles of a DFS is based on collective dephasing~\cite{Lidar_2003}. Consider a two-qubit system with a computational space spanned by basis states \(\{\ket{00},\ket{01},\ket{10}\), and \(\ket{11}\}\). Suppose this system is coupled to an environment such that the qubits undergo the dephasing process
\begin{equation}
    \ket{0}\rightarrow\ket{0}, \hspace{0.3cm}\ket{1}\rightarrow \exp{i\phi}\ket{1}
\end{equation}
for some arbitrary phase \(\phi\). Consequently, the basis states evolve as
\begin{equation}
    \begin{split}
    &\ket{00}\rightarrow\ket{00},\\
    &\ket{01}\rightarrow \exp{i\phi}\ket{01},\\
    &\ket{10}\rightarrow \exp{i\phi}\ket{10},\\
    &\ket{11}\rightarrow \exp{2i\phi}\ket{11}.
    \end{split}
\end{equation}
The states \(\ket{01}\) and \(\ket{10}\) acquire the same phase. Thus, a subspace spanned by \(\ket{0_{E}} = \ket{01}\), \(\ket{1_{E}} = \ket{10}\) expressed as an encoded state \(\ket{\psi_{E}} = a\ket{0_{E}}+b\ket{1_{E}}\) evolves according to
\begin{equation}
    \ket{\psi_{E}}\rightarrow a e^{i\phi}\ket{0_{E}}+b e^{i\phi}\ket{1_{E}} = e^{i\phi}\ket{\psi_{E}}.
\end{equation}
Global phases have no physical meaning, and, hence, \(\ket{\psi_{E}}\) remains unaffected by collective dephasing. Therefore, the space spanned by \(\{\ket{01},\ket{10}\}\) forms a DFS providing inherent robustness to collective dephasing. As exemplified, a key necessity for a DFS is the presence of symmetry amongst the basis states under the action of the system-environment couplings~\cite{Dubois2023DFSSymmetry}.

In the following, we summarize the proof of Theorem~\ref{TH:DFS} in the underlying manuscript and Theorem~4 in Reference~\cite{Lidar_2003}. The theorem is based on results by Zanardi and Rasetti~\cite{ZanardiMPLB1997ErrorAvoidingCodes} and Zanardi~\cite{ZanardiPRA1998DFS}. 

Suppose \(\{\tilde{\ket{k}}\}_{k=1}^{N}\) forms a basis for an \(N\)-dimensional subspace \(\tilde{\mathcal{H}}\) embedded in Hilbert space \(\mathcal{H}\). Consequently, a density matrix within \(\tilde{\mathcal{H}}\) follows as
\begin{equation}
    \tilde{\rho} = \sum_{k,j=1}^{N}\rho_{kj}\tilde{\ket{k}}\tilde{\bra{j}}
\end{equation}
and the action of a jump operator \(F_{\nu}\) on \(\tilde{\mathcal{H}}\) is described by
\begin{equation}
    F_{\nu}\tilde{\ket{k}} = \sum_{j=1}^{N}c^{\nu}_{kj}\tilde{\ket{j}}, \hspace{0.3cm} F_{\nu}^{\dagger}\tilde{\ket{k}} = \sum_{j=1}^{N}c^{\nu *}_{kj}\tilde{\ket{j}}.
\end{equation}
If the Lindblad equation vanishes, \(\mathcal{L}[\tilde{\rho}(t)] = 0\), \(\tilde{\mathcal{H}}\) constitutes a DFS under the action of operators \(F_\nu\). The Lindblad equation introduced in Eq.~(\ref{EQ:Lindblad}) follows as
\begin{equation}\label{EQ:LindbladGen}
\begin{array}{rcl}
    \mathcal{L}[\tilde{\rho}(t)] &=& \frac{1}{2}\sum_{\nu,\mu = 1}^{M}a_{\nu\mu}\left(
    [F_{\nu},\tilde{\rho}F^{\dagger}_{\mu}]+[F_{\nu}\tilde{\rho},F^{\dagger}_{\mu}]
    \right)\\
    &=&\frac{1}{2}\sum_{\nu,\mu=1}^{M}a_{\nu\mu}\sum_{k,j=1}^{N}\rho_{kj}\\
    &&\times\left(\sum_{n,m=1}^{N}2c^{\mu*}_{jm}c^{\nu}_{kn}\ket{\tilde{n}}\bra{\tilde{m}}\right.\\
    &&\hspace{5mm}-\sum_{n,l=1}^{N}c^{\mu*}_{nl}c^{\nu}_{kn}\tilde{\ket{l}}\tilde{\bra{j}}\\
    &&\hspace{5mm}\left.-\sum_{m,p=1}^{N}c^{\mu*}_{jm}c^{\nu}_{mp}\tilde{\ket{k}}\bra{\tilde{p}}\right)\\ 
    &=& 0.
\end{array}
\end{equation}
The terms \(a_{\nu\mu}\) represent the environment-system coupling strengths. They are unrestricted and not controllable. This implies that each summand in \(\nu,\mu\) equals \(0\). Furthermore, to derive the most general DFS condition, the result must be independent of the initial choice of \(\tilde{\rho}\). Consequently, each summand in \(k,j\) must also vanish separately. This requirement can be met if the three projection operators \(\tilde{\ket{k}}\bra{\tilde{j}}\) in Eq.~(\ref{EQ:LindbladGen}) are the same. This condition can be reflected by the choice \(c^{\nu}_{kn} = c^{\nu}_{k}\delta_{kn}\) with \(\delta_{kn}\) denoting the Kronecker delta function. Accordingly, Eq.~(\ref{EQ:LindbladGen}) simplifies to 
\begin{equation}
    \sum_{k,j=1}^{N}\rho_{kj}\tilde{\ket{k}}\tilde{\bra{j}}\left(2c^{\mu*}_{j}c^{\nu}_{k}-c^{\mu*}_{k}c^{\nu}_{k} - c^{\mu*}_{j}c^{\nu}_{j}\right)=0.
\end{equation}
For each summand to vanish, the following must hold
\begin{equation}\label{EQ:LindbladSimpl}
    \frac{c^{\mu*}_{k}}{c^{\mu*}_{j}}+\frac{c^{\nu}_{j}}{c^{\nu}_{k}} = 2.
\end{equation}
In general, Eq.~(\ref{EQ:LindbladSimpl}) is a single complex equation in multiple variables with an infinite number of solutions, and there are no further restrictions on \(c^{\nu}_{k}\). 
However, for a necessary and sufficient condition for a DFS, Eq.~(\ref{EQ:LindbladSimpl}) must also hold if \(\nu = \mu\). Imposing this restriction and substituting \(z = \frac{c^{\nu}_{j}}{c^{\nu}_{k}}\), results in
\begin{equation}
    \frac{1}{z^{*}}+z = 2
\end{equation}
with the unique solution of \(z=1\). Hence, \(c^{\nu}_{j} = c^{\nu}_{k}\), and, thus, the eigenvalues \(c^{\nu}_{k}\) are independent in \(k\). Ergo, each \(\tilde{\ket{k}}\) must be a degenerate eigenvector of operator \(F_{\nu}\).


\vspace{5mm}\textit{Appendix B: Measurement Problem \& Continuous Spontaneous Localization.---}Quantum mechanics has a multitude of interpretations. The arguably most prominent of them is the Copenhagen interpretation~\cite{Omnes1994Int}. The Copenhagen interpretation has one major drawback, the ``measurement problem'' or the ``macro-objectification problem''~\cite{heisenberg_1952_measurementproblem}. While other interpretations, such as the Many-Worlds interpretation~\cite{Everett1973MWInt}, resolve this particular issue, they typically have other drawbacks~\cite{BallentineFP1973MWInt}. Physical collapse models aim to address the measurement problem by introducing mathematical modifications to the framework underlying the conventional Copenhagen interpretation of quantum mechanics rather than an alternative interpretation. 

A measurement of a quantum system of interest---referred to as \(\mathcal{Q}\)---entails \(\mathcal{Q}\) to get entangled with a collection of quantum systems \(\mathcal{A}\). Eventually, the wave function describing the coherent evolution of the joint quantum state comprising \(\mathcal{Q}\) and \(\mathcal{A}\) collapses, and the subsequent state evolution turns incoherent. Typically, the collapse probability rises as the size of \(\mathcal{A}\) increases. 

If the interaction between \(\mathcal{Q}\) and \(\mathcal{A}\) is controlled through a measurement apparatus, the extracted classical information is subsequently available to the apparatus operator. Conversely, the extracted information is lost if the interaction between \(\mathcal{Q}\) and the environment \(\mathcal{A}\) is uncontrollable and inaccessible. In general, the collapse of the joint wave function describing \(\mathcal{Q}\) and \(\mathcal{A}\) does not necessarily lead to \(\mathcal{Q}\) being ultimately projected onto one of its eigenstates. From the point of view of \(\mathcal{Q}\), the collapse of the joint wave function is generally described as a decoherence mechanism. 

A typically undisputed axiom within the Copenhagen interpretation defines that the Schrödinger equation captures the evolution of quantum states. The Schrödinger equation is limited to describing unitary processes. Consequently, the irreversible character of measurements, particularly of quantum states in superpositions, cannot be captured by the Schrödinger equation. The linear and deterministic nature of the Schrödinger equation inherently means it cannot capture the evolution of quantum states during measurements, thus resulting in the ``measurement problem''~\cite{Bassi_2003}.

A resolution for the measurement problem is the wave packet reduction (WPR) postulate, where the superposition ``collapses'' into one of the eigenstates---quantum states with a classical representation---based on probabilities derived from the square of the inner product between the superposition state and the measured state. This collapse breaks the linearity of quantum mechanics, and the unknown boundary between quantum mechanics and classical mechanics during a measurement has puzzled theorists for decades~\cite{heisenberg_1952_measurementproblem}.

Early attempts to resolve the measurement problem include Bohm and Bub's work reconciling macroscopic observation with quantum mechanics via state vector interactions with Wiener-Siegel hidden variables~\cite{RevModPhys.38.453}. Pearle~\cite{PhysRevD.13.857}, Gisin~\cite{PhysRevLett.52.1657}, and Diosi~\cite{LDiosi_1988} introduced stochastic differential equations to model state evolution. This led to the development of the GRW theory (named in honor of its inventors Ghirardi, Rimini, and Weber and also known as Quantum Mechanics with Spontaneous Localization)~\cite{Ghirardi_PRD1986_GRW}, proposing sudden localizations for particles in a system. Despite its promise to resolve the measurement problem, the GRW theory faces challenges such as lacking a compact equation for state vector evolution and not preserving the symmetries of identical particle systems. Continuous spontaneous localization (CSL) addresses these drawbacks~\cite{Ghirardi_PRA1990_CSL}. While the original CSL framework, the model used in this work, assumes Markovian dynamics, non-Markovian versions have been developed~\cite{BassiPRA2009CSLnMarkov}. Today, CSL is the leading contender among the collapse models attempting to address the measurement problem. 

\vspace{5mm}\textit{Appendix C: Proof---Absence of Decoherence-Free Subspaces.---}For all operators \(\mathcal{N}(\textbf{x})\), evaluated at position \(\textbf{p}_1\), one can find an arbitrary number of degenerate eigenstates \(\ket{q,s}\) with eigenvalues \(n(\textbf{p}_1, q)\) such that \(n(\textbf{p}_1, q_a) = n(\textbf{p}_1, q_b)\) for two distinct sets of \(N\) indistinguishable and spatially fixed particles \(q_a=\{\textbf{q}_{a_1},\textbf{q}_{a_2}, \dots, \textbf{q}_{a_N}\}\) and \(q_b=\{\textbf{q}_{b_1},\textbf{q}_{b_2}, \dots, \textbf{q}_{b_N}\}\). 

For a DFS to exist, the same eigenstates need to remain degenerate for an operator acting on a different position \(\textbf{p}_2=\textbf{p}_1+\textbf{v}\) with \(\textbf{v}\neq\textbf{0}\), meaning, \(n(\textbf{p}_2, q_a)=n(\textbf{p}_2, q_b)\):
\begin{equation}
\begin{split}
    \mathcal{N}(\textbf{p}_2)\ket{q_a,s} &= n(\textbf{p}_2, q_a)\ket{q_a,s},\\
    \mathcal{N}(\textbf{p}_2)\ket{q_b,s} &= n(\textbf{p}_2, q_b)\ket{q_b,s}.
\end{split}
\end{equation}
Therefore, to rule out the existence of a DFS, it is sufficient to show the following: 

\begin{quote}
There does not exist a set of distinct eigenstates \(q_a\), \(q_b\) with eigenvalues\\\(n(\textbf{p}_1, q_a) = n(\textbf{p}_1, q_b)\)\\that also satisfy\\\(n(\textbf{p}_2, q_a)=n(\textbf{p}_2, q_b)\) for \(\textbf{p}_1\neq\textbf{p}_2\).
\end{quote}

\subsection{Degenerate Eigenstates at \texorpdfstring{\(\textbf{p}_1\)}{TEXT}}
Degenerate eigenstates with distinct \(q_a\) and \(q_b\) for operators \(\mathcal{N}(\textbf{x})\) acting at position \(\textbf{p}_1\) can be engineered such that \(n(\textbf{p}_1, q_a) = n(\textbf{p}_1, q_b)\). The eigenvalues \(n(\textbf{p}_1, q)\), as introduced in Eq.~(8.2) of Reference~\cite{Bassi_2003}, can be expressed after dropping the global factors \(\left(\frac{\alpha}{2\pi}\right)^{3/2}\) and \(\exp{-\frac{\alpha}{2}\vert\textbf{p}_1\vert^2}\) as
\begin{equation}
    \begin{split}
        n(\textbf{p}_1, q_a)&=\sum_i\exp{-\frac{\alpha}{2}\left(\vert\textbf{q}_{a_i}\vert^2 - 2\textbf{p}_1^T\textbf{q}_{a_i}\right)},\\
        n(\textbf{p}_1, q_b)&=\sum_i\exp{-\frac{\alpha}{2}\left(\vert\textbf{q}_{b_i}\vert^2 - 2\textbf{p}_1^T\textbf{q}_{b_i}\right)}.
    \end{split}
\end{equation}
For instance, degenerate eigenstates with eigenvalues \(n(\textbf{p}_1, q)\) can be engineered by mapping each \(\textbf{q}_{a_i}\) to a single \(\textbf{q}_{b_i}\) such that: 

\begin{equation}
    \vert\textbf{q}_{a_i}\vert^2 - 2\textbf{p}_1^T\textbf{q}_{a_i}= \vert\textbf{q}_{b_i}\vert^2 - 2\textbf{p}_1^T\textbf{q}_{b_i}.
    \label{EQ:1}
\end{equation}
Eq.~(\ref{EQ:1}) can be met if each element in \(\textbf{q}_{a_i}\) and \(\textbf{q}_{b_i}\) is either \(q_{a_{i_j}}=q_{b_{i_j}}\) or \(q_{a_{i_j}}=2p_{j}- q_{b_{i_j}}\). Therefore, it is straightforward to design degenerate eigenstates with distinct sets \(q_a\) and \(q_b\) for an arbitrary position \(\textbf{p}_1\). 

\subsection{Eigenvalues at Position \texorpdfstring{\(\textbf{p}_2\)}{TEXT}}

After defining \(\textbf{p}_2=(\textbf{p}_1+\textbf{v})\) with \(\textbf{v}\neq \textbf{0}\) and dropping global factors \(\exp{-\frac{\alpha}{2}\vert\textbf{v}\vert^2}\) and \(\exp{-\alpha \textbf{v}^{T}\textbf{p}_1}\), the eigenvalues follow as
\begin{equation}
    \begin{split}
        n(\textbf{p}_2, q_a)&=\sum_i n(\textbf{p}_1, q_{a_i})\exp{\alpha \textbf{v}^T\textbf{q}_{a_i}},\\
        n(\textbf{p}_2, q_b)&=\sum_i n(\textbf{p}_1, q_{b_i})\exp{\alpha \textbf{v}^T\textbf{q}_{b_i}}
    \end{split}
\end{equation}
with
\begin{equation}
    \begin{split}
        n(\textbf{p}_1, q_{a_i})&=\exp{-\frac{\alpha}{2}\left(\vert\textbf{q}_{a_i}\vert^2 - 2\textbf{p}_1^T\textbf{q}_{a_i}\right)},\\
        n(\textbf{p}_1, q_{b_i})&=\exp{-\frac{\alpha}{2}\left(\vert\textbf{q}_{b_i}\vert^2 - 2\textbf{p}_1^T\textbf{q}_{b_i}\right)}.
    \end{split}
\end{equation}

Representing all \(\exp{\alpha \textbf{v}^T\textbf{q}}\) as a series \(\sum_{l=0}^\infty\frac{(\alpha \textbf{v}^T\textbf{q})^l}{l!}\), one can separate the function in linearly independent contributions in the variable \(\textbf{v}\) and order \(l\):
\begin{equation}
    \begin{split}
        n(\textbf{p}_2, q_a) & = \sum_i n(\textbf{p}_1, q_{a_i})\left(1+\alpha\textbf{v}^T\textbf{q}_{a_i}+\mathcal{O}(\textbf{v}^2)\right),\\
        n(\textbf{p}_2, q_b) & = \sum_i n(\textbf{p}_1, q_{b_i})\left(1+\alpha\textbf{v}^T\textbf{q}_{b_i}+\mathcal{O}(\textbf{v}^2)\right).
    \end{split}
\end{equation}
The zeroth order in \(\textbf{v}\) is equivalent to the initial condition \(n(\textbf{p}_1, q_a)=n(\textbf{p}_1, q_b)\). After dropping the global factors \(\alpha\), the first order in \(\textbf{v}\) is expressed as

\begin{equation}\label{EQ:firstorder}
    \begin{split}
        n^{(1)}(\textbf{p}_2, q_a) & = \sum_i n(\textbf{p}_1, q_{a_i})\textbf{v}^{T}\textbf{q}_{a_i},\\
        n^{(1)}(\textbf{p}_2, q_b) & = \sum_i n(\textbf{p}_1, q_{b_i})\textbf{v}^{T}\textbf{q}_{b_i}.
    \end{split}
\end{equation}
The equation \(n(\textbf{p}_{2},q_{a}) = n(\textbf{p}_{2},q_{b})\) implies equality in all oders in \(\textbf{v}\) individually. Hence, \(n^{(1)}(\textbf{p}_2, q_a)=n^{(1)}(\textbf{p}_2, q_b)\). In the following, we will show that this assumption does not hold, and thus, the existence of a DFS can be ruled out.


Suppose, in the most general case, there is a weighted sum in \(f(\textbf{x}, \textbf{q}_{b_i})=\exp{-\frac{\alpha}{2}(\vert\textbf{q}_{b_i}\vert^2-2\textbf{x}^T\textbf{q}_{b_i})}\) to represent each \(f(\textbf{x}, \textbf{q}_{a_i})=\exp{-\frac{\alpha}{2}(\vert\textbf{q}_{a_i}\vert^2-2\textbf{x}^T\textbf{q}_{a_i})}\):
\begin{equation}
    \begin{split}
        f(\textbf{p}_1, \textbf{q}_{a_i}) &= \sum_j w_{i,j}f(\textbf{p}_1, \textbf{q}_{b_j})\\
         e^{-\frac{\alpha}{2}(\vert\textbf{q}_{a_i}\vert^2 -2\textbf{p}_1^T\textbf{q}_{a_i})} &= \sum_j w_{i,j}  e^{-\frac{\alpha}{2}(\vert\textbf{q}_{b_j}\vert^2-2\textbf{p}_1^T\textbf{q}_{b_j})}
    \end{split}
    \label{EQ:expand}
\end{equation}
with \(\sum_i w_{i,j} = 1\).
Then, according to Eq.~(\ref{EQ:firstorder}), the first order in \(\textbf{v}\) follows as
\begin{equation}      
    \textbf{v}^{T}\!\textbf{q}_{a_i} \!e^{\!-\!\frac{\alpha}{2}\!\vert\textbf{q}_{a_i}\!\vert^2 +\alpha\textbf{p}_1^T\!\textbf{q}_{a_i}} \!\!= \!\!\sum_j\! w_{i,j} \textbf{v}^{T}\!\textbf{q}_{b_j} \!e^{\!-\!\frac{\alpha}{2}\!\vert\textbf{q}_{b_j}\!\vert^2+\alpha\textbf{p}_1^T\!\textbf{q}_{b_j}}\!.
    \label{EQ:O1cond1}
\end{equation}
Alternatively, using Eq.~(\ref{EQ:expand}), the first order in \(\textbf{v}\) can be written as
\begin{equation}   
    \textbf{v}^{T}\!\textbf{q}_{a_i}\! e^{\!-\!\frac{\alpha}{2}\!\vert\textbf{q}_{a_i}\!\vert^2 \!+\alpha\textbf{p}_1^T\!\textbf{q}_{a_i}} \!\!=\! \!\textbf{v}^{T}\!\textbf{q}_{a_i}\!\sum_j\! w_{i,j}  e^{\!-\!\frac{\alpha}{2}\!\vert\textbf{q}_{b_j}\!\vert^2\!+\alpha\textbf{p}_1^T\!\textbf{q}_{b_j}}\!.
    \label{EQ:O1cond2}
\end{equation}

Given that the functions \(f(\textbf{x},\textbf{q})=\exp{-\frac{\alpha}{2}(\vert\textbf{q}\vert^2-2\textbf{x}^T\textbf{q})}\) are linearly independent for differing \(\textbf{q}\) 
it follows that Eqs.~(\ref{EQ:O1cond1}) and (\ref{EQ:O1cond2}) can only hold simultaneously if \(\textbf{q}_{a_i}=\textbf{q}_{b_j}\) and there is only a single \(w_{i,j}\neq0\) leading to
\begin{equation}        
    \textbf{v}^{T}\textbf{q}_{a_i} e^{-\frac{\alpha}{2}\vert\textbf{q}_{a_i}\vert^2 +\alpha\textbf{p}_1^T\textbf{q}_{a_i}} = \textbf{v}^{T}\textbf{q}_{b_j} e^{-\frac{\alpha}{2}\vert\textbf{q}_{b_j}\vert^2+\alpha\textbf{p}_1^T\textbf{q}_{b_j}}.
\end{equation}
Therefore, the sets \(q_a\) and \(q_b\) are equivalent. 

Consequently, for \(\textbf{v}\neq\textbf{0}\), \(n^{(1)}(\textbf{p}_2, q_a)=n^{(1)}(\textbf{p}_2, q_b)\), and, thus, \(n(\textbf{p}_2, q_a)=n( \textbf{p}_2, q_b)\), is true if and only if \(q_a=q_b\).

\vspace{5mm}
In conclusion, eigenstates with two distinct sets \(q_a\) and \(q_b\) can only be degenerate for operators \(\mathcal{N}(\textbf{x})\) acting on a single position \(\textbf{p}_1\). Consequently, a decoherence-free subspace cannot be engineered for systems governed by continuous spontaneous localization using Gaussian functions and, more generally, radial basis functions. 


Note, non-degenerate eigenvectors cannot form an eigenbasis. Therefore, linear combinations of non-degenerate eigenvectors at position \(\textbf{p}_2\) cannot form a ``logical'' eigenbasis. Hence, encoding a linear combination of eigenvectors at position \(\textbf{p}_2\) cannot achieve a degeneracy, and, thus, form a DFS.

\vspace{5mm}\textit{Appendix D: Continuous Spontaneous Localization---Mass-Density Operator.---}A conceptual extension of the CSL model, driven by the quest for a relativistic version and a more accurate alignment with quantum observations~\cite{Bassi_2003}, is the replacement of the number-density operator \(\mathcal{N}(\textbf{x})\) in Eq.~(\ref{number density op}) with a mass-density operator
\begin{equation}
    M(\textbf{x}) = \sum_{k}\frac{m_{k}}{m_{0}}N_{k}(\textbf{x}).
\end{equation}
Here, \(m_{k}\) denotes the masses of particles of species \(k\), \(m_{0}\) is a chosen reference mass typically corresponding to the nucleon mass, and \(N_{k}(\textbf{x})\) signifies the local number density operator \(\mathcal{N}(\textbf{x})\) for particles of species \(k\). This formulation enables a more nuanced characterization of the system. 

To proceed, creation and annihilation operators need additional sub-indices specifying particle type: \(a^{\dagger}_{k}(\textbf{y}_{k},s_{k}), a_{k}(\textbf{y}_{k},s_{k})\). These operators adhere to the commutation relation
\begin{equation}
    [a_{k}(\textbf{y}_{k},s_{k}),a^{\dagger}_{k'}(\textbf{y}_{k'}',s_{k'}')]_{\pm} = \delta(\textbf{y}_{k'}'-\textbf{y}_{k})\delta_{s_{k},s_{k'}'}\delta_{k,k'},
\end{equation}
where \([\cdot]_{\pm}\) denotes the anti-commutator or commutator for fermionic and bosonic excitations, respectively. If the subscript sign is not specified, the commutator is implied. The complete mass density operator \(M(\textbf{x})\) becomes
\begin{equation}
    M\!(\textbf{x})\! =\!\! \sum_{k}\!\frac{m_{k}}{m_{0}}\!\sum_{s_{k}}\!\!\int\! \!d\textbf{y}_{\!k} g(\textbf{y}_{\!k}\!\!-\!\textbf{x}) a^{\dagger}_{k}(\textbf{y}_{\!k},\!s_{k})a_{k}(\textbf{y}_{\!k},\!s_{k}).
\end{equation}

The probability measure for the reduction process \(g(\textbf{x})\) remains unchanged, as it solely depends on the position and should not differentiate between different particles. Consequently, the resultant Lindblad equation, Eq.~(\ref{EQ:CSL}), now reads
\begin{equation}
    \begin{array}{rcl}
        \mathcal{L}[\rho(t)] &=& \frac{\gamma}{2m_{0}^2}\left(\int d\textbf{x}[M(\textbf{x}),\rho(t)M(\textbf{x})]\right.\\
             &&\left.+\int d\textbf{x}[M(\textbf{x})\rho(t),M(\textbf{x})]\right),
    \end{array}
\end{equation}
where \(\gamma\) remains unaltered from the mass-independent version in Eq.~(\ref{EQ:CSL}). The eigenvectors in Eq.~(\ref{EQ:Eigen}) are generalized as
\begin{equation}
    \ket{q,s} = \mathfrak{N}a^{\dagger}_{k_{1}}(\textbf{q}_{k_{1}},s_{k_{1}})
    \dots
    a^{\dagger}_{k_{n}}(\textbf{q}_{k_{n}},s_{k_{n}})\ket{0}
\end{equation}
with corresponding eigenvalues
\begin{equation}
    \begin{split}
    n(\textbf{x}) &= \sum_{j=1}^{n}\frac{m_{k_{j}}}{m_{0}}g(\textbf{q}_{k_{j}}-\textbf{x})\\
    &= \left(\frac{\alpha}{2\pi}\right)^{3/2}\sum_{i=j}^{n}\frac{m_{k_{j}}}{m_{0}}e^{-\frac{\alpha}{2}(\textbf{q}_{k_{j}}-\textbf{x})^{2}},
    \end{split}
\end{equation}
where the subscript \(k_{j}\) identifies the \(j\)'th excitation with the particle species \(k_{j}\) and corresponding mass \(m_{k_{j}}\). The resulting condition for a DFS is that all states \(\ket{q_i,s}\) within the DFS are degenerate with respect to \(M(\textbf{x})\)
\begin{equation}
    \begin{split}
    M(\textbf{x})\ket{q_{i},s} &= \left(\frac{\alpha}{2\pi}\right)^{\frac{3}{2}}\sum_{j=1}^{n}\frac{m_{k_{j}}}{m_{0}}e^{-\frac{\alpha}{2}(\textbf{q}_{k_{j},i}-\textbf{x})^{2}}\ket{q_{i},s}
    \end{split}
\end{equation}
for all \(q_{i}\) and for all \(\textbf{x}\). Notably, the resulting eigenvalues for the mass-density CSL closely resemble the mass-independent CSL eigenvalues, differing only by constant prefactors \(\frac{m_{k_{j}}}{m_{0}}\). However, as demonstrated in Appendix C, 
Gaussian functions are linearly independent, making it impossible to construct two distinct eigenstates \(\ket{q_{a},s},\ket{q_{b},s}\) that are degenerate with respect to \(M(\textbf{p}_{1})\) \textit{and} \(M(\textbf{p}_{2})\) for \(\textbf{p}_{1}\neq \textbf{p}_{2}\). Thus, a DFS is unattainable in this scenario as well.

\begin{figure*}[t]
    \includegraphics[width=\textwidth]{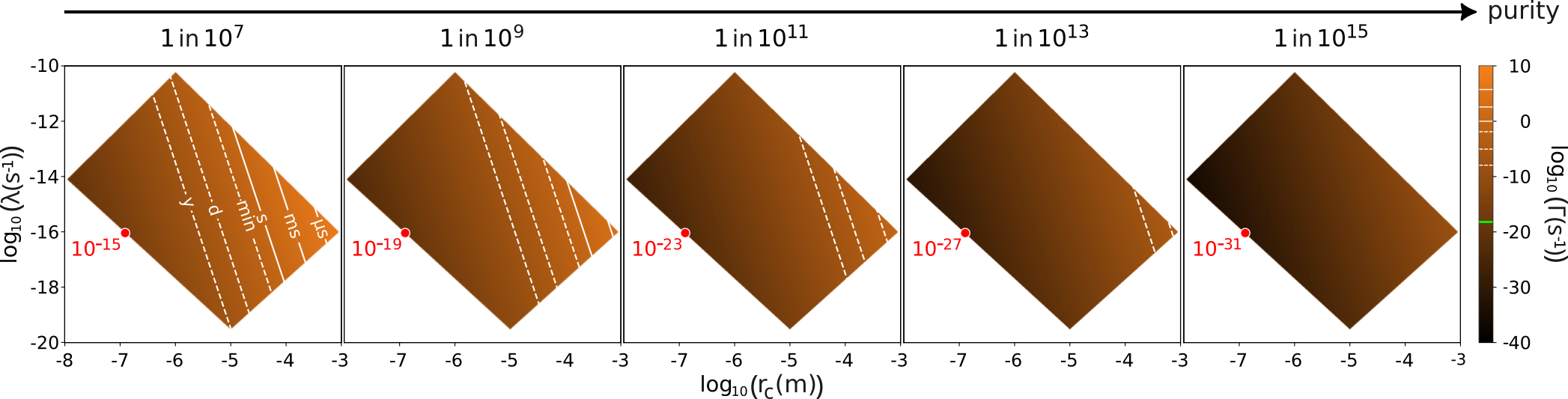}
    \caption{\textbf{Unexplored CSL Parameter Range and Resulting Wave Function Collapse Rates.} \(\lambda\) represents the microscopic decay rate, while \(r_c\) delineates the boundary between quantum and classical regimes. \(\Gamma\) denotes the anticipated macroscopic collapse rate (\(\Gamma = \lambda * n^2\)), with \(n\) representing the number of constituent interacting quantum systems (e.g., two-level systems) within a volume \(r_c^3\). The subpanels' titles indicate the density of quantum systems interacting directly and indirectly with a qubit. The analysis here is conducted for CSL and white noise. The white areas may be excluded as potential parameter spaces, as noted in prior literature~\cite{CarlessoNatPhys2022CollapseReview}. Nonetheless, these parameter ranges remain relevant, as the CSL noise mechanism is anticipated to exhibit colored characteristics. The dashed lines represent coherence limits exceeding one second, while solid lines denote coherence limits up to one second, as labeled in the leftmost panel. In the case of silicon as a host material---one of the most pristine commercially available materials---the central panel indicates the highest commercially-feasible purity. The collapse rate resulting from the parameter suggestions made in the original GRW Reference~\cite{Ghirardi_PRD1986_GRW} is shown in red. The green line on the color bar corresponds to the age of the universe.}
    \label{Fig_App}
\end{figure*}

\vspace{5mm}\textit{Appendix E: Continuous Spontaneous Localization---Parameters.---}In a composite system with a superposition of states separated by at least \(r_{c}\equiv 1/\sqrt{\alpha}\), following Reference~\cite{Bassi_2014}, the collapse rate of the system's center of mass can be estimated using
\begin{equation}
    \Gamma = \lambda n^{2} N.
\end{equation}
In this context, \(\lambda\) represents the collapse rate of an individual quantum constituent, \(n\) indicates the number of constituents within a volume of radius \(r_{c}\), and \(N\) signifies the total number of such volumes in the system. Fig.~\ref{Fig_App} depicts the collapse rate \(\Gamma\) for various densities \(n\) of two-level systems (TLS) due to impurities and crystal defects within a block of silicon (\(N=1\)). These TLSs may become entangled with nearby qubits, causing the qubit's wave function to spread, thereby exacerbating decoherence.

Silicon stands out for its exceptional purity, making it a premier platform for quantum processors, including superconducting and semiconductor qubits. However, defects in silicon significantly impact processor performance, spanning from surface irregularities like pits to deeper imperfections within the bulk. These defects encompass vacancies, interstitials, dislocations, stacking faults, and anomalies introduced during crystal growth. Despite efforts to mitigate them, some defects may persist, potentially compromising wafer quality.

Commercially available high-purity silicon wafers typically have an impurity density of less than one impurity in \num{e9} atoms, with the highest purity levels reaching around one in \num{e11}, albeit available only in polycrystalline form~\cite{Tokuyama}. As materials undergo processing, defect and impurity densities typically increase, particularly near surfaces and interfaces, sometimes by several orders of magnitude~\cite{Mitchell_2023}.

Not all defects and impurities can form quantum systems interacting with qubits within their operational frequency range. Those that do play a critical role in quantum information processing, influencing phenomena such as decoherence and relaxation~\cite{MüllerIOP2019TLS,GraafScieAdv2020TLSQuasi,ThorbeckPRXQ2023TLSIonization}. These TLSs can interact and form clusters that increase wave function spread.

Recent studies have highlighted the severity of coherence limits induced by solid-state materials, with losses exceeding previous estimations by one order of magnitude~\cite{ChecchinPRApp2022Si}. Additionally, doping in silicon, particularly with boron acceptors, can lead to strongly coupled TLS baths for superconducting qubits~\cite{zhang2024acceptorinduced}.

Estimating the number of TLS within silicon reveals intriguing insights. For undoped high-purity silicon, the intrinsic charge carrier density is around \SI{e16}{carriers\per\meter\cubed}, resulting in an impurity density of around one in \num{e12}, in agreement with achieved purities~\cite{Tokuyama}. TLS density estimates for superconduting qubits range around \SI{e20}{\per\meter\cubed\per\GHz}~\cite{MartinisPRL2005TLSDensity,BarendsPRL2013TLSDensity}, leading to an impurity ratio of one in \num{e8} for a \SI{1}{\GHz} range. Therefore, the range of interacting TLS densities for silicon or other less-pristine materials can be expected to be at best between \num{e7} and \num{e15}, as illustrated in Fig.~\ref{Fig_App}.

The coherence time of naturally or implanted impurities in ultra-pristine and isotopically purified silicon can achieve coherence times of several tens of minutes~\cite{SaeediScience2013T1}. However, the coherence time of artificial qubits, such as superconducting qubits~\cite{SomoroffPRL2023T2, WangNPJQ2022T2} or semiconductor qubits~\cite{HansenAPR2022T2, ZhouNatPhys2024T2}, typically falls within the range of single-digit milliseconds. This stark contrast in coherence times underscores the significant dependence on material quality. The coherence times observed, alongside the estimated range of collapse rates in Fig.~\ref{Fig_App}, suggest that the coherence of artificial qubits may be approaching or have already reached this fundamental limit.

\vspace{5mm}
It is worth noting that all discussions here center on white-noise CSL models. However, white noise exhibits a spectrum that is uniformly flat and independent of frequency, lacking a clear physical basis. A more plausible approach involves noise from tangible sources, necessitating a colored spectrum with non-trivial characteristics. References~\cite{AdlerIOP2007colorednoise, AdlerIOP2008colorednoise, Carlesso_Ferialdi_Bassi_2018} delve into the concept of colored noise, delineating cut-off points and exploring the potential cosmological origins of noise. They model it as a scalar field, and the calculated thermal correlation functions are then mapped to physical sources or particles. This endeavor aims to bridge the gap between theoretical models and empirical observations. Consequently, this discussion underscores various generalizations of the CSL model that strive for enhanced physical validity, introducing a novel domain of reduction parameters awaiting experimental validation.

\bibliography{references.bib}
\end{document}